\documentclass[aps,pra,amsmath,amssymb,twocolumn,superscriptaddress,showpacs,floatfix]{revtex4-1}

\usepackage{bm}
\usepackage{epsfig}
\usepackage[usenames]{color}
\usepackage{soul}
\usepackage{graphicx}
\usepackage[hypertex]{hyperref}
\usepackage{srctex}

\newcommand{\bareline}{
\frac{\quad }{\quad} {\!}{\!}  \frac{\quad }{\quad} {\!}{\!}
\frac{\quad}{\quad} }

\newcommand{\updownline}{
\frac{\quad }{\quad} {\!}{\!}{\!}{\!} \uparrow {\!}{\!}{\!}{\!}
\frac{\;\;\:}{\;\;\;} {\!}{\!}{\!}{\!} \downarrow {\!}{\!}{\!}{\!}
\frac{\quad}{\quad} {\!}{\!} \frac{\;\:}{\;\;}}

\newcommand{\uplineleft}{
\frac{\quad }{\quad} {\!}{\!}{\!}{\!} \uparrow {\!}{\!}{\!}{\!}
 {\!}{\!}{\!}{\!} \frac{\qquad}{\qquad} {\!}{\!} \frac{\;\;}{\;\;}}

\newcommand{\uplineright}{
\frac{\quad }{\quad} {\!}{\!}{\!}{\!} \frac{\;\;\;}{\;\;\;}
{\!}{\!} \uparrow {\!}{\!}{\!}{\!} {\!}{\!}{\!}{\!}
\frac{\qquad}{\qquad} }

\newcommand{\downlineleft}{
\frac{\quad }{\quad} {\!}{\!}{\!}{\!} \downarrow {\!}{\!}{\!}{\!}
 {\!}{\!}{\!}{\!} \frac{\qquad}{\qquad} {\!}{\!} \frac{\;\;}{\;\;}}

\newcommand{\downlineright}{
\frac{\quad }{\quad} {\!}{\!}{\!}{\!} \frac{\;\;\;}{\;\;\;}
{\!}{\!} \downarrow {\!}{\!}{\!}{\!} {\!}{\!}{\!}{\!}
\frac{\qquad}{\qquad} }

 \DeclareMathOperator{\Real}{Re}
\DeclareMathOperator{\Imag}{Im}

\DeclareMathOperator{\TUH}{\mathrm{\scriptscriptstyle TUJ}}
\DeclareMathOperator{\KK}{\mathrm{\scriptscriptstyle KK}}
\DeclareMathOperator{\JH}{\mathit J_{\mathrm{\scriptscriptstyle
H}}} \DeclareMathOperator{\HT}{\mathit
H_{\mathrm{\scriptscriptstyle T}}}
\DeclareMathOperator{\HTS}{\mathit H^2_{\mathrm{\scriptscriptstyle
T}}} \DeclareMathOperator{\HU}{\mathit
H_{\mathrm{\scriptscriptstyle U}}}
\DeclareMathOperator{\HJ}{\mathit H_{\mathrm{\scriptscriptstyle
J}}}

\renewcommand{\paragraph}[1]{\textit{#1.---} } 

\begin{document}

\date{\today }

\title{Effective orbital ordering in multiwell optical lattices with fermionic atoms }

\author{A.~M.~Belemuk}
\affiliation{Institute for High Pressure Physics, Russian Academy of Science, Troitsk 142190, Russia}
\affiliation{Department of Theoretical Physics, Moscow Institute of Physics and Technology, 141700 Moscow, Russia}

\author{N.~M.~Chtchelkatchev}
\affiliation{Department of Theoretical Physics, Moscow Institute of Physics and Technology, 141700 Moscow, Russia}
\affiliation{Department of Physics and Astronomy, California State University Northridge, Northridge, CA 91330, USA}
\affiliation{L.D. Landau Institute for Theoretical Physics, Russian Academy of Sciences,117940 Moscow, Russia}

\author{A.~V.~Mikheyenkov}
\affiliation{Institute for High Pressure Physics, Russian Academy of Science, Troitsk 142190, Russia}
\affiliation{Department of Theoretical Physics, Moscow Institute of Physics and Technology, 141700 Moscow, Russia}

\date{\today}

\begin{abstract}
We consider the behavior of Fermi atoms on optical superlattices
with two-well structure of each node. Fermions on such lattices
serve as an analog simulator of Fermi type Hamiltonian. We derive
a mapping between fermion quantum ordering in the optical
superlattices and the spin-orbital physics developed for
degenerate $d$-electron compounds. The appropriate effective
spin-orbital model appears to be the modification of the
Kugel-Khomskii Hamiltonian. We show how different ground states of
this Hamiltonian correspond to particular spin-pseudospin
arrangement patterns of fermions on the lattice. The dependence of
fermion arrangement on phases of complex hopping amplitudes is
illustrated.
\end{abstract}


\maketitle

\section{Introduction}\label{sec.intro}
Experimental investigations of ultracold atoms in optical lattices
have opened up a unique flexibly tunable simulator for study of
quantum many-body
physics~\cite{Greiner2002Nature,Kohl2005PRL,kawaguchi2012spinor,hauke2012}
in the parameter range that had been hardly possible or even
impossible to achieve in the natural solid state
systems~\cite{Meacher1998,Bruder1998PRL,Ovchinnikov2000,Corboz2013PRX}.

Atom temperature on the optical lattice can be made extremely
low. It opens the experimental way to
investigate in detail the structure of the ground state and the
low-lying many-body states of atoms~\cite{Bloch08_RMP,Paul2012ChemPhys}.
One of the most interesting
regimes corresponds to the strong atom-atom quantum correlations.
Interactions between atoms on the lattice have different nature.
Atoms can jump (tunnel) from site to site of the optical lattice
with the characteristic hoping energy $t$. Within the site
typically there is repulsion $U$ between atoms. While the atoms
have spins there is exchange interaction between the spins of the
atoms on the neighboring sites of the lattice. The quantum state
of the atoms on the lattice also strongly depends on the
statistics of atoms, either they are bosons or
fermions~\cite{Lewens07_AP}. In what follows we shall focus on the
fermion case.

Typically atoms on the lattice could be well described by
modifications of the Hubbard model due to the short-range
character of the atom-atom interaction $U$~\cite{Bruder1998PRL}.
The problem of the ground state and the low-lying many-body states
of atoms on the lattice have been successfully investigated within
the mean-field theory, see e.g. Ref.~\onlinecite{Lewens07_AP}.
Progress have also been made beyond the mean-field theory in
particular with numerical simulations of the Hubbard-type models.
For bosons on the lattice parameter range of $U$ and $t$ at which
one could expect Bose-condensation or the Mott-insulator behavior
was thoroughly investigated~\cite{Lewens07_AP, Bruder2011PRA}.
Experimental realization of a Mott insulator regime of fermions on
the optical lattice~\cite{Jordens08} opened a unique possibility
to simulate various ground states and spin orderings of fermions,
complying with theoretical predictions for the repulsive
Fermi-Hubbard model.

Recently optical lattices with complicated structure of the node
attracted much attention, in particular, superlattices with
two-well structure~\cite{anderlini2007,
folling2007Nature,anderlini2007Nature,trotzky2008Science,Wagner2012PRA}.
The mean-field ground-state phase diagram of spinor bosons in
two-well superlattice was found using Bose-Hubbard Hamiltonian in
Ref.~\onlinecite{Wagner2012PRA}. It was shown that the system
supports Mott-insulating as well as superfluid phases like in
one-well latices. But the quadratic Zeeman effect lifts the
degeneracy between different polar superfluid phases leading to
additional metastable phases and first-order phase transitions.

Here we focus our study on spinor fermions on optical
superlattices with multi-well structure of each node.
Specifically, we consider two-well nodes in the regime of strong
correlations (large $U/t$). We show how the ground many-body atom
state on the lattice can be understood without direct solving of
the Hubbard model but using the well known results of the
machinery developed long ago for degenerate $d$-electron
compounds~\cite{kugel1973JETP,Kugel1982UFN}. We show that there is
a mapping between fermion quantum ordering in the optical
superlattices and the spin-orbital physics of degenerate
$d$-electron compounds. We derive the effective spin-orbital model
and show that it appears to be the generalization of the
Kugel-Khomskii Hamiltonian~\cite{kugel1973JETP}. Different ground
states of this Hamiltonian correspond to particular nontrivial
fermion arrangement on the lattice.

The paper is organized as follows: In the beginning of
Sec.~\ref{sec:model} we write down the Hubbard-type Hamiltonian
for fermions on multi-well lattice. Then in Sec.~\ref{subsec:EH}
more or less standard steps have been done to reduce the model to
the effective spin-orbital Hamiltonian. Some rather cumbersome
technical details of the reduction we put in the Appendix. In the
Discussions, Sec.~\ref{secDiscussion}, we give examples of
possible atom many-body ground states on the lattice that can be
obtained from the mapping to orbital-spin physics.

\section{Microscopic model for the fermions in the double-well optical lattice \label{sec:model}}

\subsection{Tunnel Hamiltonian model\label{subsec:TH}}
\begin{figure}[h]
  \centering
  \includegraphics[width=0.7\columnwidth]{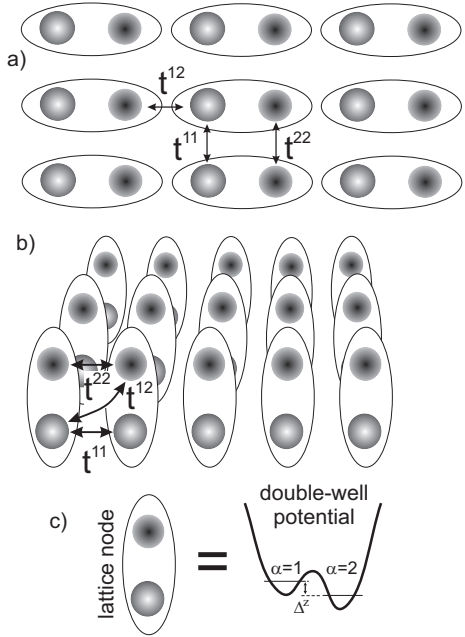} \\
  \caption{a,b) The sketch of possible optical lattices with two-well structure
  where spin-orbital effects may show up. $t^{\alpha \beta}$ are hopping amplitudes between
  wells on nearest nodes. Indices $\alpha, \beta= 1, 2$ numerate the two wells at a given lattice node
  (two quantum pseudospin states). c) The structure of the lattice node. $\Delta^z$ is the
  energy offset between the two wells. }\label{fig1}
\end{figure}

We consider the $d$-dimensional hypercube optical lattice where
each node is a double well, as is illustrated for the two-dimensional
lattice in Fig.~\ref{fig1}.

The Hamiltonian describing the quantum states of fermions on the
lattice can be written as
\begin{gather}\label{eqH}
  H=H_{\Delta}+\HT+\HU +\HJ.
\end{gather}
The term $H_\Delta$ describes the level structure of each node
\begin{gather}
  H_\Delta=\sum_{i, \sigma,\alpha,\beta} \frac12 \left( \Delta_i^{z}\sigma^z_{\alpha\beta}+
  \Delta_i^{x}\sigma^x_{\alpha\beta} \right) {\,} c^\dag_{i  \alpha \sigma} c_{i \beta \sigma },
\end{gather}
where index $i$ labels the nodes, $\alpha=1,2$ is the well number
at a given node, $\Delta_i^{z}$ is the difference of the ground
state energies between the two wells, while $\Delta_i^{x}$ takes
into account possible tunneling between the wells in a node.
$\sigma^z$ and $\sigma^x$ are Pauli matrices. Operator
$c^{\dag}_{i \alpha \sigma }(c_{i \alpha \sigma })$ is the fermion
creation (annihilation) operator for fermion atom residing at a
node $i$, in a well $\alpha$ with spin projection~$\sigma$.

Tunneling between the nodes specifies
\begin{gather}
  \HT= -\sum_{i\neq j,\sigma,\alpha,\beta} t_{ij}^{\alpha\beta} c^\dag_{i\alpha\sigma}c_{j\beta\sigma},
\end{gather}
where $t_{ij}^{\alpha\beta}$ is the tunnel matrix element. The
structure of the tunnel matrix elements is schematically
depicted in Fig.~\ref{fig1}a,b. The hopping amplitudes
$t_{ij}^{\alpha \beta}$ can be arranged into complex-valued
amplitude matrix in the well-space:
\begin{equation}
t_{ij}^{\alpha \beta}= t^{\alpha \beta}=
\begin{pmatrix}
t^{11} &t^{12}
\\
t^{21} &t^{22}
\end{pmatrix}.
\end{equation}
We shall omit for brevity the lattice indices in hopping
amplitudes. Below notation $t^\dag= t^\dag_{ij}$ will be used for
the Hermitian conjugation in the well-subspace. Note that, in
general, $t\neq t^\dag$. Due to the Hermitian character of $\HT$
there is a standard symmetry, $t_{ij}^{\alpha
\beta}=(t_{ji}^{\beta \alpha})^*$. It follows that $t^\dag$
corresponds to the hopping amplitude matrix with interchanged
lattice indices, i.e. $(t^\dag)^{\alpha \beta}= (t_{ij}^{\beta
\alpha})^*= t_{ji}^{\alpha \beta}$.

Since each node has the ``fine'' structure related to the wells it
is convenient to split the interaction Hamiltonian into two parts,
$\HU+ \HJ$. The first term has a trivial structure in the well
index space, and describes the Coulomb repulsion $(U_i > 0)$ of
fermions at one node:
\begin{gather}
 \HU=\sum_{i,\sigma,\sigma',\alpha,\alpha'}U_i n_{i\alpha\sigma}
 n_{i,\alpha',\sigma'}(1-\delta_{\alpha\alpha'}\delta_{\sigma\sigma'}),
\end{gather}
where $n_{i\alpha\sigma}= c^\dag_{i\alpha\sigma} c_{i\alpha
\sigma}$. The second term describes the ferromagnetic Hund's
coupling~\cite{Kugel1982UFN} $(\JH^{(i)} > 0)$ between fermions in
wells $\alpha= 1$ and 2 at a given lattice node
\begin{gather}
  \HJ=-\sum_{i,\sigma,\sigma'}\JH^{(i)} c_{i,1,\sigma}^\dag c_{i, 1,\sigma'} c_{i,2,\sigma'}^\dag c_{i,2,\sigma},
\end{gather}
This term comes into effect if the average fermion density at a
node $\langle n_i \rangle= \sum_{\sigma} (\langle n_{i1\sigma}
\rangle + \langle n_{i2\sigma} \rangle)$ is equal to $\langle n_i
\rangle= 2$.

\subsection{The effective Hamiltonian for single-atom filling of the nodes\label{subsec:EH}}

We shall focus on the case when $U_i$ is the largest energy scale,
in particular $U_i$ is much larger than the hopping amplitudes,
$t_{ij}^{\alpha\beta}$. Then each node, on average, is occupied by
one fermion and the Hamiltonian~\eqref{eqH} can
be simplified. To proceed, we introduce standard
presentation~\cite{abrikosov1965pseudospin} of the spin $S= 1/2$
and the pseudospin $\tau= 1/2$ operators through the fermion
creation and annihilation operators, see, e.g.,
Ref.~\onlinecite{Yamashita1998PRB}:
\begin{gather}\label{eqSc}
  S^a_i= \frac12 c^\dag_{i\alpha \sigma} \sigma^a_{\sigma\sigma'} c_{i\alpha \sigma'},
  \\\label{eqtauc}
  \tau^a_i= \frac12 c^\dag_{i\alpha \sigma} \sigma^a_{\alpha\beta} c_{i\beta \sigma}.
\end{gather}
Index $a= x ,y, z$, or sometimes, it is convenient to use $a= 1,
2, 3$. Summation over recurring spin and pseudospin indices is
implied. We remind that representation~\eqref{eqSc}-\eqref{eqtauc}
is valid only at the single-atom filling of each node.

Below we focus on the case when the interactions, $U$ and $\JH$,
do not depend on the site index. Using \eqref{eqSc} and
\eqref{eqtauc} we can present the term $H_\Delta$ in the form
$H_\Delta=\sum_i\left(\Delta_i^{z} \tau^z_i+\Delta_i^x
\tau^x_i\right)$. The term $H_{\TUH}=\HT+\HU +\HJ$ after the
standard perturbation procedure in hopping
amplitudes~\cite{kugel1973JETP,Kugel1982UFN, Auerbach1994book,
Feiner1997PRL,Ishihara1997PRB, Yamashita1998PRB,Brzezicki2013PRB}
can be transformed into the following general form [derivation
details we put in Appendix~\ref{Ap1}]
\begin{multline}\label{HKH}
  H_{\TUH}=\sum_{\langle i,j\rangle} \biggl[ \frac14 A_{ij}+ A_{ij} \mathbf S_i\cdot \mathbf S_j+
  B_{ij}^{a b} \tau_i^a \tau_j^b- \frac12 K^a_{ij}{\,} (\tau_i^{a}+ \\ \tau_j^{a})+
  4 {\,} \mathbf S_i\cdot \mathbf S_j {\,} \Bigl \{ D_{ij}^{ab} \tau_i^{a} \tau_j^b+
  \frac12 K^a_{ij} {\,} (\tau_i^{a}+ \tau_j^{a}) \Bigr \} \biggr],
\end{multline}
where the summation runs over bonds $\langle i,j\rangle$ between
nearest neighbors. Coefficients $A_{ij}$ $B^{ab}_{ij}$,
$K^a_{ij}$, and $D^{ab}_{ij}$ are quadratic in the tunnel
amplitudes $t^{\alpha \beta}_{ij}$ and can be considered as
generalized exchange coupling constants of the resulting spin-spin,
spin-pseudospin and pseudospin-pseudospin interactions
between fermions. Vectors $K^a_{ij}$ introduce as well an
effective magnetic field into the pseudospin space, resulting from
nondiagonal structure of the hopping matrix $t^{\alpha \beta}$.

For particular case of real hopping amplitudes, $t^{11}= t^{22}=
t$, $t^{12}= t^{21}= 0$ and zero Hund's coupling $J_H= 0$, the
model \eqref{HKH} is equivalent to the Hamiltonian of the $SU(4)$
model~\cite{Yamashita1998PRB}
\begin{equation}
H_{\TUH}\to \frac{2t^2}{U} {\,} \sum_{\langle i,j\rangle} \left( \frac12 +
2 \mathbf S_i\cdot \mathbf S_j \right) \left( \frac12 + 2 {\bm
\tau}_i \cdot {\bm \tau}_j \right)
\end{equation}

If we identify the space of well indices with the ``orbital''
space then the Hamiltonian~\eqref{HKH} for real
$t_{ij}^{\alpha\beta}$ becomes similar to the Kugel-Homsky
Hamiltonian~\cite{kugel1973JETP} developed for degenerate
$d$-electron compounds.\cite{Kugel1982UFN, Auerbach1994book,
Feiner1997PRL,Ishihara1997PRB,Yamashita1998PRB,Brzezicki2013PRB}

Eq.~\eqref{HKH} has been derived assuming $\JH/U\ll1$. However
in $d$-electron compounds it is quite often that $\JH \sim U$. In
a similar way it may take place for atoms on the optical lattice.
The conjecture has been made in Ref.~\onlinecite{Kugel1982UFN}
that the form of interaction terms the Kugel-Khomskii Hamiltonian
remains the same for $\JH \sim U$ and tensor coefficients $A$,
$K$, $B$ and $D$ would preserve their symmetry structure in the
orbital space. For the case of diagonal hopping amplitude matrix
$t^{\alpha\beta} \sim \delta^{\alpha\beta}$ this conjecture has
been confirmed in Ref.~\onlinecite{Kugel1982UFN} by direct
calculation of the Kugel-Khomskii Hamiltonian coefficients in all
orders in $\JH/U$. The same conclusion applies for atoms on the
lattice described by the effective Hamiltonian~\eqref{HKH}.

\section{Discussion\label{secDiscussion}}

\subsection{Symmetrical Hamiltonian. \label{subsec:SymCase}}

Now we focus on the symmetrical case when the nearest neighbor
hopping matrix $t^{\alpha \beta}$ is diagonal in the orbital
space. This case could be realized in the optical lattice sketched
in Fig.~\ref{fig1}b. Then $K^a_{ij}$ is equal to zero while
$B^{ab}_{ij}$ and $D^{ab}_{ij}$ are diagonal matrices in the
orbital space. For this case the symmetrical model Hamiltonian
follows from Eq.~\eqref{HKH} (see Appendix~\ref{Ap1})
\begin{multline}\label{HKHs}
H_{\TUH}\to  H_{\rm sym}= \sum_{\langle i j \rangle}\left\{J_1\,{\mathbf S}_i\cdot {\mathbf S}_j+\right.
\\
\left.J_2\,{\boldsymbol \tau}_i\cdot {\boldsymbol
\tau}_j+4J_3({\mathbf S}_i\cdot {\mathbf S}_j)\,({\boldsymbol
\tau}_i\cdot {\boldsymbol \tau}_j)\right\},
\end{multline}
where we shall consider exchange constants $J_{1}$, $J_2$ and $J_3$
as independent input parameters.

\begin{figure}
  \centering
  \includegraphics[width=1.0\columnwidth]{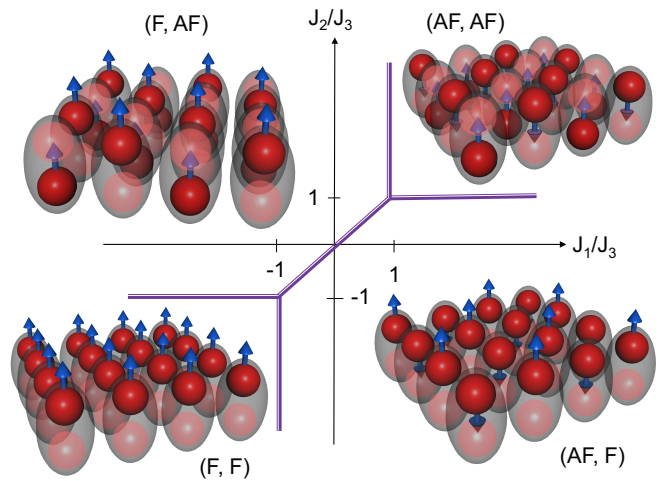}\\
  \caption{(Color online) The mean field phase diagram of symmetrical
  model~\eqref{HKHs} for $J_3>0$, see, e.g., Refs. \cite{Oles2000PRB,ishihara1996PhysicaC} for $d$-electron compounds.
  Here ``F'' stands for ferromagnetic ordering, and ``AF'' is for antiferromagnetic ordering.
  The first and the second abbreviation in the
  designations of phases are for the spin and pseudospin subsystems, respectively.
  The ordering patterns of atoms on the optical lattice
  are shown in the insets. For ferromagnetic orbital arrangement atoms
  are localized in one particular type of sub-wells (for example in the upper sub-wells).
  For antiferromagnetic orbital arrangement atoms alternate
  between the lower and upper sub-wells. Red spheres show the lattice site with the maximum
  probability of occupation by atom, while transparent spheres show ``nearly'' empty sites.
  Arrows indicate spin directions.} \label{fig4}
\end{figure}

Let us consider the most interesting case $\Delta^{x,z}\ll
J_{1,2,3}$ and we can neglect the term $H_\Delta$ comparing with
$H_{\TUH}$. Then the isolated minima of the double-well potential
are the same. The spin-pseudospin interaction resulted from
virtual hoppings between neighboring notches gives rise to the
occupancy of that sub-well which is more preferable.

The properties of the Kugel-Khomskii symmetrical
Hamiltonian~\eqref{HKHs} have been well investigated, see e.g.,
Refs.~\onlinecite{kugel1973JETP, Kugel1982UFN}. In Fig. \ref{fig4}
we present the result of the analysis of the model~\eqref{HKHs} in
the mean-field approximation for $J_3>0$ [similarly would look like the figure for $J_3<0$~\cite{Kugel1998PRB}]. The figure shows
possible phases of spin-pseudospin arrangements for various values
of exchange parameters. For example, the case $J_1>J_3>J_2>0$
corresponds to the ground state of $H_{\rm sym}$ which is
antiferromagnetic in the spin space and ferromagnetic in the
pseudospin space, (AF-F) phase in Fig.~\ref{fig4}. The effective
orbital exchange can be estimated as $J_\tau^{\rm
eff}=J_2+4J_3\langle {\mathbf S}_i\cdot {\mathbf S}_j\rangle$.
Similarly the effective spin-exchange is approximately equal to
$J_s^{\rm eff}=J_1+4J_3\langle {\boldsymbol
\tau}_i\cdot{\boldsymbol \tau}_j\rangle$. When spins are
antiferromagnetically ordered $J_\tau^{\rm eff}=J_2-J_3 < 0$ and
one obtains ``orbital ferromagnetism''. If we turn on the external
effective magnetic field we can change the orbital ferromagnetism
to orbital antiferromagnetism when the field is sufficiently
strong that $\langle {\textbf S}_i\cdot {\textbf
S}_j\rangle>J_2/4J_3$. Finite $\Delta^{x}$, $\Delta^{z}$ play the
role of the built-in effective magnetic field in the pseudo-spin
space. Large enough $\Delta^{z}$ would also drive the system into
the ferromagnetic orbital state (in such a case one of the two
minima of the double well is much lower than the other).

To illustrate the possible arrangement patterns of atoms in real
space let us consider pseudospin (orbital) state in the mean field
approximation . It can be presented as a product of one-site
orbital states, $|\psi_{\rm mf} \rangle= \prod_i |\theta_i
\varphi_i \rangle$. The orbital one-site state $|\theta_i
\varphi_i \rangle$ can be chosen as
\begin{equation} |\theta_i \varphi_i \rangle=
\cos \frac{\theta_i}{2} |1 \rangle+ e^{i\varphi_i} \sin
\frac{\theta_i}{2} |2 \rangle.
\end{equation}
The direction (in pseudospin space) of the averaged pseudospin
$\langle {\bm \tau}_i \rangle$ is defined in terms of the polar
and azimuth angles
\begin{equation}
\langle \theta_i \varphi_i | {\bm \tau}_i |\theta_i \varphi_i
\rangle= \frac12 \left( \sin \theta_i \cos \varphi_i, {\:} \sin
\theta_i \sin \varphi_i, {\:} \cos \theta_i \right).
\end{equation}
The orbital state
\begin{equation}
|\pi- \theta_i, {\;} \pi+ \varphi_i \rangle= \sin
\frac{\theta_i}{2} |1 \rangle- e^{i\varphi_i} \cos
\frac{\theta_i}{2} |2 \rangle
\end{equation}
is orthogonal to $|\theta_i {\,} \varphi_i \rangle$ and sets
$\langle {\bm \tau}_i \rangle$ in the opposite direction.
Ferromagnetic orbital arrangement corresponds to identical orbital
states $|\theta_i {\,} \varphi_i \rangle= |\theta {\,} \varphi
\rangle$ at different sites. Antiferromagnetic orbital state
corresponds to $|\theta_i {\,} \varphi_i \rangle= |\theta {\,}
\varphi \rangle$ at sublattice $i \in A$, and $|\theta_j {\,}
\varphi_j \rangle= |\pi- \theta, {\,} \pi+ \varphi \rangle$ at
sublattice $i \in B$. The average pseudospin vectors alternate at
the sublattices $A$ and $B$, $\langle {\bm \tau}_i \rangle=
-\langle {\bm \tau}_j \rangle$.

The most simple illustration of the orbital arrangement of atoms
can be given for the case of $\theta= 0$, or $\theta= \pi$. Then
atoms with probability equal to one occupy either well $\alpha=
1$, or $\alpha =2$, respectively. The illustrative example of
``phase diagrams'' for this case is sketched in Fig.~\ref{fig4}
where we adopted results of Refs.~\onlinecite{Feiner1997PRL,
Kugel1998PRB,Oles2000PRB,ishihara1996PhysicaC} on the
Kugel-Khomskii model to our problem of atom arrangements on the
optical lattice (see also Supplementary
Material~\onlinecite{Supplemental}). The sketch shows the ordering
patterns of atoms on the optical lattice of the type presented in
Fig.~\ref{fig1}b. For the ferromagnetic orbital arrangement atoms
are localized in one of the sub-wells, for example in upper
sub-wells. For the antiferromagnetic arrangement atoms alternates
between $\alpha= 1$ and $\alpha= 2$ wells (upper and lower wells
in figure). If we consider the antiferromagnetic orbital
arrangement beyond the mean field approximation then atoms are
spread between two sub-wells with some probability due to quantum
fluctuation. Red spheres in Fig.~\ref{fig4} show lattice sites
with the maximum probability of occupation by atom, while white
spheres show ``nearly'' empty sites. Arrows indicate spin
directions. The phase boundaries in Fig.~\ref{fig4} actually do
not exactly match coordinate axes in $(J_1,J_2)$ space: the
absolute value and sign of $J_3$ specify the position of the phase
boundaries~\cite{Feiner1997PRL, Kugel1998PRB,Oles2000PRB,
ishihara1996PhysicaC}, as is illustrated.

\subsection{Complex hopping amplitudes}

One of the unique properties of optical lattices is the
possibility to tune the complex tunnel amplitudes by manipulating
the laser field~\cite{Struck2012PRL}. It includes also the
possibility to manipulate the Hamiltonian by changing the phases
of the hopping amplitudes $t_{ij}$ and leaving their absolute
values fixed (i.e. no geometric distortion of the optical
lattice).

\paragraph{Toy model}
To illustrate the importance of the complex phases of the hopping
amplitudes $t_{ij}$ we consider the following toy-model. We suppose that $\JH= 0$ and we account for those hoppings which go
through different orbitals (wells):
\begin{equation}
t^{11}= 0, \quad t^{22}= 0, \quad t^{12}= t', \quad t^{21}=t'e^{i\chi}.
\end{equation}
The constant phase $\chi$ accounts for phase difference in the
non-diagonal hopping amplitudes. Then the effective Hamiltonian
\eqref{HKH} can be written as (see Appendix~\ref{Ap1})
\begin{multline} \label{Heff4}
H_{\chi}= J {\,} \sum \limits_{\langle i j \rangle}
\Bigl(\frac12+2{\,}{\bf S}_i \cdot {\bf S}_j \Bigr) \Bigl(\frac12+
2\cos\chi {\,}(\tau^x_i \tau^x_j- \tau^y_i \tau^y_j)+ \\
+ 2\sin \chi {\,} (\tau^x_i \tau^y_j+ \tau^y_i \tau^x_j)-2\tau^z_i
\tau^z_j \Bigr).
\end{multline}

The appearance of the phase-dependent ground state can be
illustrated as the following. For the ferromagnetic spin
background the mean field energy $E_{\rm mf}= \langle \psi_{\rm
mf}|H_{\chi} |\psi_{\rm mf} \rangle$ of pseudospin sub-system is
\begin{multline} \label{Emfl}
E_{\rm mf}= \frac{J}{2} \sum \limits_{\langle i,j \rangle}
\Bigl[1+ \cos (\chi- (\varphi_i + \varphi_j)) \sin \theta_i
\sin\theta_j -
\\
-\cos \theta_i \cos \theta_j \Bigr],
\end{multline}
where we used $1/2+ 2 {\bf S}_i {\bf S}_j= 1$. Consider now the
energy of the antiferromagnetic orbital state. For such a state
the mean field energy per site is
\begin{equation} \label{Emf}
E_{\rm mf}= \frac{3}{2}J \Bigl[1- \cos (\chi- 2\varphi) \sin^2 \theta + \cos^2 \theta \Bigr].
\end{equation}
The minimization of the energy $E_{mf}$ relative to $\theta$ and
$\varphi$ gives the twofold degenerate ground state $E_{gs}=
0$ with $\theta= \pi/2$ and $\varphi= \chi/2, \chi/2+ \pi$, 
The resulting direction of the pseudospin $\langle {\bm \tau}
\rangle$ depends on the phase $\chi$. In real space this state
describes the situation when the atoms with equal probability are
spread over the first and second wells in the notch but the phase
relation between pseudospin states $|1 \rangle$ and $|2 \rangle$
are tuned by the applied phase $\chi$. The change of $\chi$
induces the corresponding variation of the phase $\varphi$, which
is equivalent to rotation of the pseudospin vector $\langle {\bm
\tau} \rangle$ in the pseudospin space.

\section*{Conclusions}
Optical lattices are quantum simulators of many-particle systems.
We have shown that there is a mapping between fermion quantum
ordering in the optical superlattices and the spin-orbital physics
developed for degenerate $d$-electron compounds. The effective
spin-pseudospin model has been derived. This model is the
generalization of the Kugel-Khomskii Hamiltonian for complex
hopping amplitudes. We have shown how different ground states of
this Hamiltonian correspond to particular nontrivial fermion
arrangements on the lattice.

\acknowledgments

The work was funded by RFBR, NSF Grant DMR 1158666, the Grant of
President of Russian Federation for support of Leading Scientific
Schools, RAS presidium and Russian Federal Government programs.

\appendix
\section{Perturbative expansion in hopping amplitudes\label{Ap1}}

In the subspace of functions $|\Phi^0 \rangle$ with occupancy
equal to one at each site the hopping term $\HT$ creates
intermediate states with double occupancy. There are six different
intermediate states with double occupancy at a given site $i$,
which differ in the well $\alpha $ and spin $\sigma$ indices
\begin{equation} \label{psi1}
|\psi_1 \rangle =
\begin{pmatrix}
\bareline \\ \updownline
\end{pmatrix}, \qquad |\psi_2 \rangle =
\begin{pmatrix}
\updownline \\
\bareline
\end{pmatrix}
\end{equation}
\begin{equation} \label{psi3}
|\psi_3 \rangle =
\begin{pmatrix}
\uplineright \\ \uplineleft
\end{pmatrix}, \qquad |\psi_4 \rangle =
\begin{pmatrix}
\downlineright \\
\downlineleft
\end{pmatrix}
\end{equation}
\begin{equation} \label{psi5}
|\psi_5 \rangle =
\begin{pmatrix}
\downlineright \\ \uplineleft
\end{pmatrix}, \qquad |\psi_6 \rangle =
\begin{pmatrix}
\uplineright \\ \downlineleft
\end{pmatrix}
\end{equation}
Here the lower (upper) level is for the pseudospin state $\alpha=
1 (2)$. All of them are eigenstates of the $\HU$ with the same
energy $U$ and the first four are also eigenstates of $\HJ$.
Although the term $\HJ$ mixes the states $|\psi_5 \rangle$ and
$|\psi_6 \rangle$ it mixes them into eigenstate of $\HU$.

In the second order perturbation theory in hopping term $\HT$ the
effective Hamiltonian has the form~\cite{Auerbach1994book}
\begin{equation}
H_{\TUH}= -\HT \frac{1}{\HU+ \HJ} \HT.
\end{equation}
Assuming that $\JH \ll U$ the above expression in first order of
$\JH/U$ can be simplified as
\begin{equation}
H_{\TUH}= -\HT \left[ \frac{1}{\HU}- \frac{1}{\HU} \HJ.
\frac{1}{\HU} \right] \HT
\end{equation}
As we mentioned above all the intermediate states \eqref{psi1},
\eqref{psi3} and \eqref{psi5} after mixing them by $\HJ$ remain
eigenstates of $\HU$. It enables to reduce the above expression
for $H_{\TUH}$ to the following form
\begin{equation} \label{Heff1}
H_{\TUH}= -\frac1U \left(\HTS-  \frac{1}{U} \HT \HJ \HT\right).
\end{equation}
Presentation of fermi-operators through the spin and pseudospin
operators which is originally due to Kugel and
Khomskii~\cite{kugel1973JETP} can be given as
\begin{equation}
c^{\dagger}_{i \alpha \gamma} c_{i \beta \gamma'}= \left( \frac12
\delta_{\alpha \beta}+ \tau_i^{a} \sigma^a_{\beta \alpha} \right)
\left( \frac12 \delta_{\gamma \gamma'}+ S_i^{b} \sigma^b_{\gamma'
\gamma} \right).
\end{equation}

In the subspace of functions $|\Phi^0 \rangle$ the first and the
second term of the $\HTS$ is reduced to
\begin{widetext}
\begin{multline} \label{HTS1}
\HTS= \sum \limits_{\langle i j \rangle} \biggl \{ {\rm Sp}
(t^{\dagger} t)+ {\rm Sp}(t^{\dagger} \sigma^a t) {\,}\tau^a_i+ {\rm Sp}(t \sigma^a t^{\dagger}) {\,}\tau^a_j-
\\
\left( \frac12 + 2 {\,} {\bf S}_i \cdot {\bf S}_j \right) \Bigl[
\frac12 {\rm Sp}(t^{\dagger} t)+ {\rm Sp}(t^{\dagger} \sigma^a t)
{\,} \tau^a_i+ {\rm Sp}(t \sigma^a t^{\dagger}){\,} \tau^a_j+ 2
{\,} {\rm Sp}(t^{\dagger} \sigma^a t\sigma^b t) {\,} \tau^a_i
\tau^b_j \Bigr] \biggr \},
\end{multline}
\begin{multline} \label{HTS2}
\frac{1}{U} \HT \HJ \HT= \left( \frac{-\JH}{U} \right) \sum
\limits_{\langle i j \rangle} \biggl \{ -\frac12 {\,} {\rm
Sp}(t^{\dagger} \sigma^{a'}t) \tau^{a'}_i- \frac12 {\,} {\rm Sp}
(t \sigma^{a'} t^{\dagger}) \tau^{a'}_j- {\rm Sp}(t^{\dagger}
\sigma^{a'} \sigma^b t) \tau^{a'}_i \tau^{b}_j- {\rm
Sp}(t^{\dagger} \sigma^b t \sigma^{a'}) \tau^b_i
\tau^{a'}_j+
\\
\left( \frac12 + 2 {\bf S}_i \cdot {\bf S}_j \right) \Bigl[
\frac12 {\,} {\rm Sp}(t^{\dagger} t)+ \frac12 {\,} {\rm
Sp}(t^{\dagger} \sigma^{a'} t) \tau^{a'}_i+ \frac12 {\,}{\rm Sp}(t
\sigma^{a'} t^{\dagger}) \tau^{a'}_j- {\rm Sp}(t^{\dagger}
\sigma^z t\sigma^a ) \tau^z_i \tau^a_j- {\rm Sp}(t^{\dagger}
\sigma^a t \sigma^z ) \tau^a_i \tau^z_j \Bigr] \biggr \}
\end{multline}
The summation over repeated indices $a, b= 1, 2, 3$ is implied,
indices with prime mean that the summation does not include the
third component, i.e. $a', b'= 1, 2$. In terms proportional
to $\tau^a_j$ we used the property of the Hermitian conjugate
hopping matrix $t^{\alpha \beta}_{ji}= (t^\dag)^{\alpha
\beta}_{ij}$.

In what follows we omit the constant term ${\rm Sp} (t^{\dagger}
t)$ in $\HTS$. Gathering both terms together we obtain
after regrouping the following effective Hamiltonian
\begin{multline} \label{HKHa}
H_{\TUH}= \sum \limits_{\langle i j \rangle} {\,} \biggl[ \frac14
A_{ij} + A_{ij} {\,} {\bf S}_i \cdot {\bf S}_j + B^{ab}_{ij} {\,}
\tau^a_i \tau^b_j -\frac12 ( K_{ij}^a {\,} \tau^a_i+ K_{ji}^a {\,}
\tau^a_j)+ 4 {\:} {\bf S}_i \cdot {\bf S}_j \Bigl \{ D^{ab}_{ij}
{\,} \tau^a_i \tau^b_j+ \frac12 (K_{ij}^a {\,} \tau^a_i+ K_{ji}^a
{\,} \tau^a_j) \Bigr \} \biggr],
\end{multline}
\end{widetext}
where
\begin{gather}\label{eqA}
A_{ij}= \frac{1}{U} {\,} {\rm Sp}(t\,t^\dag) \left(1-
\frac{\JH}{U}
\right). \\
B^{ab}_{ij}= \frac{1}{U} {\,} {\rm Sp}(t^\dag \sigma^a t \sigma^b)
\left\{
\begin{aligned}
1&+ \frac{2 \JH}{U},    &a,b= 1,2 \\
1&+ \frac{3 \JH}{2U},   &{a= 1,2, b= 3 \atop a= 3, b= 1,2} \\
1&+ \frac{\JH}{U},      &a,b= 3 \\
\end{aligned}
\right. \\
D^{ab}_{ij}= \frac{1}{U} {\,} {\rm Sp}( t^\dag \sigma^a t
\sigma^b) \left\{
\begin{aligned}
1&,                 &a,b= 1,2 \\
1&+ \frac{\JH}{2U}, &{a= 1,2, b= 3 \atop a= 3, b= 1,2} \\
1&+ \frac{\JH}{U},  &a,b= 3 \\
\end{aligned}
\right. \\
K^{a}_{ij}= \frac{1}{U} {\,}{\rm Sp}(t^\dag \sigma^a t) \left\{
\begin{aligned}
1&- \frac{\JH}{2U},  &a&= 1,2 \\
1&,                  &a&= 3 \\
\end{aligned}
\right.
\end{gather}
Vectors $K^a_{ji}$, which enter in Eq. \ref{HKHa}, are
proportional to ${\rm Sp}(t^\dag_{ji} \sigma^a t_{ji})$. They can
be given in terms of $t_{ij}$ using the equality ${\rm
Sp}(t^\dag_{ji} \sigma^a t_{ji})= {\rm Sp}(t_{ij} \sigma^a
t^\dag_{ij})$. Note also that for zero Hund's coupling, $J_H= 0$,
the second rank tensors are the similar, $B^{ab}_{ij}=
D^{ab}_{ij}$.

The presentation in the form \eqref{HKHa} can be viewed as a
generalization of the corresponding
Kugel-Khomskii~\cite{kugel1973JETP} Hamiltonian for complex
hopping amplitudes. Below we write down the explicit form of all
the traces that contribute to the coefficients of the Hamiltonian:
\begin{eqnarray*}
{\rm Sp}(t^{\dagger} t)&=& |t^{11}|^2+ |t^{22}|^2+ |t^{12}|^2+ |t^{21}|^2 \\
{\rm Sp}(t^{\dagger} \sigma^z t)&=& |t^{11}|^2- |t^{22}|^2+ |t^{12}|^2- |t^{21}|^2 \\
{\rm Sp}(t \sigma^z t^{\dagger})&=& |t^{11}|^2- |t^{22}|^2+
|t^{21}|^2- |t^{12}|^2 \\
{\rm Sp}(t^{\dagger}\sigma^z t \sigma^z)&=& |t^{11}|^2+
|t^{22}|^2- |t^{12}|^2- |t^{21}|^2 \\
{\rm Sp}(t^{\dagger} \sigma^x t)&=& 2 \Real[t^{11}(t^{21})^*+ t^{22}(t^{12})^*] \\
{\rm Sp}(t \sigma^x t^{\dagger})&=& 2\Real[t^{11}(t^{12})^*+ t^{22}(t^{21})^*] \\
{\rm Sp}(t^{\dagger} \sigma^y t)&=& 2\Imag[-t^{11}(t^{21})^*+
t^{22}(t^{12})^*] \\
{\rm Sp}(t \sigma^y t^{\dagger})&=& 2\Imag[t^{11}(t^{12})^*-
t^{22}(t^{21})^*] \\
{\rm Sp}(t^{\dagger}\sigma^x t \sigma^x)&=& 2\Real[t^{11}(t^{22})^*+
t^{12}(t^{21})^*] \\
{\rm Sp}(t^{\dagger}\sigma^y t \sigma^y)&=& 2\Real[t^{11}(t^{22})^*-
t^{12}(t^{21})^*] \\
{\rm Sp}(t^{\dagger} \sigma^x t \sigma^y)&=& 2\Imag[t^{11}(t^{22})^*- t^{12}(t^{21})^*] \\
{\rm Sp}(t^{\dagger} \sigma^y t \sigma^x)&=&
2\Imag[-t^{11}(t^{22})^*- t^{12}(t^{21})^*] \\
{\rm Sp}(t^{\dagger}\sigma^x t \sigma^z)&=& 2\Real[t^{11}(t^{21})^*-
t^{22}(t^{12})^*] \\
{\rm Sp}(t^{\dagger}\sigma^z t \sigma^x)&=& 2\Real[t^{11}(t^{12})^*-
t^{22}(t^{21})^*] \\
{\rm Sp}(t^{\dagger} \sigma^z t \sigma^y)&=& 2\Imag[t^{11}(t^{12})^*+
t^{22}(t^{21})^*] \\
{\rm Sp}(t^{\dagger} \sigma^y t \sigma^z)&=&
2\Imag[-t^{11}(t^{21})^*- t^{22}(t^{12})^*]
\end{eqnarray*}
For specific choices of $t^{\alpha \beta}$, in particular for
those considered in the paper, $K^a_{ji}= K^a_{ij}$, and the
general form \eqref{HKHa} is reduced to the Eq. \ref{HKH} of the
main text.

For the case of real site-independent hopping amplitudes $t^{11}=
t_1$, $t^{22}= t_2$, $t^{12}= t^{21}= t_{12}$ the Hamiltonian
\eqref{HKHa} is reduced to the original Kugel-Khomskii Hamiltonian
\begin{widetext}
\begin{multline} \label{HKHb}
H_{\KK}= \frac1U \sum \limits_{\langle i j \rangle} \biggl[
-(t_1^2- t_2^2)(\tau^z_i+ \tau^z_j)- \left(1- \frac{J_H}{2U}
\right) 2t_{12}(t_1+ t_2) (\tau^x_i+ \tau^x_j)+
\\
\frac{J_H}{U} {\,} 4(t_1t_2+ t_{12}^2) \tau^x_i \tau^x_j+
\frac{J_H}{U} {\,} 4(t_1t_2- t_{12}^2) \tau^y_i \tau^y_j+
\frac{J_H}{U} {\,}
2t_{12}(t_1- t_2)(\tau^z_i \tau^x_j+ \tau^x_i \tau^z_j) +
\\
\Bigl( \frac12 + 2 {\bf S}_i \cdot {\bf S}_j \Bigr) \biggl\{
\left(1- \frac{J_H}{U} \right) \frac12 (t_1^2+ t_2^2+ 2t_{12}^2)+
(t_1^2- t_2^2) (\tau^z_i+ \tau^z_j)+ \left(1 -\frac{J_H}{2U}
\right) 2t_{12}(t_1+ t_2) (\tau^x_i+ \tau^x_j)+
\\
4(t_1t_2+ t_{12}^2) \tau^x_i \tau^x_j+ 4(t_1t_2- t_{12}^2)
\tau^y_i \tau^y_j+ \left( 1+ \frac{J_H}{U} \right) 2(t_1^2+ t_2^2-
2t_{12}^2) \tau^z_i \tau^z_j+ \left(1+ \frac{J_H}{2U} \right)
4t_{12}(t_1- t_2)(\tau^z_i \tau^x_j+ \tau^x_i \tau^z_j) \biggr\}
\biggr]
\end{multline}
For diagonal hopping matrix $t_{12}= 0$, and $t_1= t_2= t$ the
Hamiltonian \eqref{HKHb} is simplified to
\begin{gather} \label{HKHc}
H_{\KK}= \frac14 J_1+ J_1 {\bf S}_i \cdot {\bf S}_j+ J_2 {\bm
\tau}_i \cdot {\bm \tau}_j+ 4 J_3 ({\bf S}_i \cdot {\bf S}_j) {\,}
({\bm \tau}_i \cdot {\bm \tau}_j)- J_3 \Bigl( 1- 4{\,} {\bf S}_i \cdot {\bf S}_j \Bigr)
\frac{J_H}{U}{\,} \tau^z_i \tau^z_j
\end{gather}\end{widetext}
where
\begin{gather}
J_1= \frac{2t^2}{U} \Bigl(1- \frac{J_H}{U} \Bigr), \\ J_2=
\frac{2t^2}{U} \Bigl(1+ 2\frac{J_H}{U} \Bigr), \\ J_3=
\frac{2t^2}{U}
\end{gather}
The Hamiltonian serves as a starting form for the symmetrical
Hamiltonian \eqref{HKHs} with independent $J_1$, $J_2$ and $J_3$,
if one neglects anisotropy term ($\propto J_H/U {\,} \tau^z_i
\tau^z_j$) in pseudospin space.

\paragraph{Toy model} To illustrate the meaning of complex phases we have considered hereinabove the case
when hopping process can be described by the following toy-model
\begin{equation}
t^{11}= 0, \quad t^{22}= 0, \quad t^{12}= t', \quad
t^{21}=t'e^{i\chi}.
\end{equation}
For such amplitudes the only nonzero traces are
\begin{gather}
{\rm Sp}(t t^{\dagger})= 2|t'|^2, \quad
{\rm Sp}(\sigma^z t \sigma^z t^{\dagger})= -2|t'|^2 \\
{\rm Sp}(\sigma^x t \sigma^x t^{\dagger})= -{\rm Sp}(\sigma^y t \sigma^y t^{\dagger})= 2|t'|^2 \cos \chi \\
{\rm Sp}(\sigma^x t \sigma^y t^{\dagger})= {\rm Sp}(\sigma^y t
\sigma^x t^{\dagger})= 2|t'|^2 \sin \chi
\end{gather}

The Hamiltonian $H_{\TUH}$ can be rewritten as
\begin{equation} \label{Hchi}
H_{\chi}= J {\,} \sum \limits_{\langle i j \rangle} \Bigl(\frac12+
2{\,}{\bf S}_i \cdot {\bf S}_j \Bigr) \Bigl( \frac12 + 2 B^{ab}
{\,} \tau^a_i \tau^b_j \Bigr)
\end{equation}
where effective exchange is $J= 2|t'|^2/U$ and $B^{12}= B^{21}=
\sin \chi$, $B^{11}= -B^{22}= \cos \chi$, $B^{33}= -1$. The Eq.
\ref{Hchi} can be rewritten in the form \eqref{Heff4}.

\bibliography{edu_bib}
\end{document}